\documentstyle[aas2pp4,natbib209]{article}

\begin{document}

\title{Macroscopic Dust in Protoplanetary Disks - From Growth to Destruction}

\author{J. Deckers\altaffilmark{1} and J. Teiser}
\affil{Fakult\"at f\"ur Physik, Universit\"at Duisburg-Essen, 47057 Duisburg}

\altaffiltext{1}{johannes.deckers@uni-due.de}
\begin{abstract}
The collision dynamics of dusty bodies are crucial for planetesimal formation. Decimeter agglomerates are especially important in the different formation models. Therefore, in continuation of our experiments on mutual decimeter collisions, we investigate collisions of centimeter onto decimeter dust agglomerates in a small drop tower under vacuum conditions ($p\lesssim5\cdot10^{-1}~\mathrm{mbar}$) at a mean collision velocity of $6.68\pm0.67~\mathrm{m\,s^{-1}}$. We use quartz dust with irregularly shaped micrometer grains. Centimeter projectiles with different diameters, masses and heights are used, their typical volume filling factor is $\Phi_{p,m}$$=0.466\pm0.02$. The decimeter agglomerates have a mass of about 1.5\,kg, a diameter and height of 12\,cm and a mean filling factor of $\Phi_{t,m}$$=0.44\pm0.004$. At lower collision energies only the projectile gets destroyed and mass is transferred to the target. The accretion efficiency decreases with increasing obliquity and increasing difference in filling factor, if the projectile is more compact than the target. The accretion efficiency increases with increasing collision energy for collision energies under a certain threshold. Beyond this threshold at $298\pm25~\mathrm{mJ}$ catastrophic disruption of the target can be observed. This corresponds to a critical fragmentation strength $Q^*=190\pm16~\mathrm{mJ\,kg^{-1}}$, which is a factor of four larger than expected. Analyses of the projectile fragments show a power law size distribution with average exponent of $-3.8\pm0.3$. The mass distributions suggest that the fraction of smallest fragments increases for higher collision energies. This is interesting for impacts of small particle on large target bodies within protoplanetary disks, as smaller fragments couple better to the surrounding gas and re-accretion by gas drag is more likely.
\end{abstract}

\section{Introduction}
Planets form in disks of gas and dust around young stars, the protoplanetary disks. It is widely accepted that the process of planet formation starts with micron-sized dust grains. Evidence for this is found by astronomical observations \citep{pinte2008, hernandez2007} and by analysis of meteorites \citep{brearley1999, scott2005}. Out of these grains the km-sized planetesimals are formed, which can grow to ever larger bodies by accretion and will eventually form planets. The formation processes involved in the growth of these planetesimals however are not understood in detail, yet.\\
Various models try to explain the formation of planetesimals. These can roughly be divided into two main groups. One group of models is based on growth through mutual collisions between dust agglomerates. Collisions of grains can lead to sticking and bigger aggregates are formed. In this way aggregates can grow efficiently to millimeter sizes \citep{blum2008, zsom2010, windmark2012a}. However, the bigger the agglomerates get the more problematic growth by mutual collisions becomes. At aggregate sizes in the millimeter range mutual collisions do not necessarily lead to sticking, but rather lead to bouncing \citep{guettler2010, jankowski2012}. In case of larger particles also fragmentation of aggregates is a typical collision result \citep{teiser2011a, schraepler2012}. On the other hand, in collisions of aggregates of different sizes the bigger body can gain mass and grow even at higher collision velocities \citep{teiser2011b, kothe2010, meisner2013, guettler2010}. Part of the mass sticks to the bigger body directly. In addition to that, experiments \citep{wurm2001b, wurm2001a} as well as simulations \citep{jankowski2014} show that small particles ejected after the impact couple well to the surrounding gas and can be re-accreted onto the larger body by gas drag. These studies are conducted in free molecular flow, where the mean free path of the surrounding gas is big compared to the target, or in the transition regime to viscous laminar flow. \citet{sellentin2013} analyzed the collisions of small particles onto large targets in numerical simulations and found that the ejecta are not re-accreted assuming viscous laminar flow, where the mean free path is small in comparison to the target. Numerical simulations showed that growth can be efficient even in a regime where bouncing is dominant, as a broad velocity and size distribution can provide few fast and larger particles, which can sweep up smaller particles \citep{windmark2012b, windmark2012a}. Theoretical collision studies also predict that agglomerates of grains in the 100\,nm range can grow to much larger sizes than agglomerates of micron-sized grains, especially if they consist not only of silicates but of water ice \citep{okuzumi2012, kataoka2013}. If particles can grow to such large sizes and stay highly porous as predicted, then bouncing will start to dominate at much larger aggregate sizes. Growth could be very efficient then especially in the outer parts of protoplanetary disks, as the bouncing barrier could be shifted beyond the critical size range where the lifetimes of agglomerates are extremely short due to radial drift. However, currently there is no experimental proof for agglomerates in this parameter range.\\
The second group of models considers concentration of particles to high densities and subsequently a gravitation forced collapse. Gravitational attraction is not important for the interaction of (individual) small particles, as their masses are still small. Concentrating particles to high densities can result in gravitational instability as a dense cloud of particles can collapse under its own weight. Several different mechanisms to achieve these high particle densities are being discussed. These include particle concentration by baroclinic instability \citep{lyra2011}, turbulence \citep{johansen2006, lambrechts2012} or streaming instability \citep{johansen2007, youdin2005}. \citet{chiang2010} give a review of different models.\\
Decimeter bodies and their collision dynamics play an important role in both groups of models. They are of interest for the coagulation models as bodies start to decouple from the surrounding gas and drift inward toward the star when they reach the decimeter to meter range. It is therefore crucial to concentrate these particles and form larger bodies before they drift into the star and are lost for planetesimal formation. This also illustrates the high relevance of decimeter bodies for the models taking gravitational attraction into account, as the concentration mechanisms are strongest for particles on the verge of decoupling from the gas. Here, the collision dynamics of the decimeter bodies are interesting as well. Even in areas of high particle concentration mutual collisions can result in fragmentation and the generation of small particles. These small particles might not stay in the area of high particle concentration.\\
The outcome of a collision can be described by $\mu ={M_f}/{M_0}$, the ratio of the masses of the largest fragment $M_f$ and the original body $M_0$. For planetesimal formation the threshold condition between bouncing ($\mu$=1) and fragmentation ($\mu<$1) as well as the condition for catastrophic disruption ($\mu$=0.5) are of importance. These threshold conditions have been studied by \citet{beitz2011} and by \citet{schraepler2012} for collisions of centimeter sized spheres and cylinders and by \citet{deckers2013} for mutual collisions of cylindrical decimeter bodies. The specific energy $Q$, defined as ratio of threshold collision energy and mass of the collision partners, makes the results for collisions at different sizes comparable. For collisions of mutual decimeter agglomerates $Q_{\mu=1}$ is at $5 \cdot 10^{-3} \mathrm{J\,kg^{-1}}$ \citep{deckers2013}. In this study we expanded these experiments in order to investigate the threshold to catastrophic disruption $Q_{\mu=0.5}$, commonly referred to as critical fragmentation strength $Q^*$, for decimeter dust agglomerates. As the threshold conditions can not be analyzed in mutual collisions, because agglomerates are too fragile for high accelerations, we analyzed collisions of centimeter projectiles onto decimeter targets.\\

\section{Experiment}
Collision experiments are conducted under vacuum conditions in a small drop tower with a height of about 3\,m (experimental setup see Fig. \ref{aufbau}). The mean collision velocity is $6.68~\mathrm{m\,s^{-1}}$. It follows the distribution in Fig. \ref{gvert} with a standard deviation of $0.67~\mathrm{m\,s^{-1}}$.\\

\subsection{Sample Preparation}

\begin{figure}[ht]
\epsscale{0.8}
\plotone{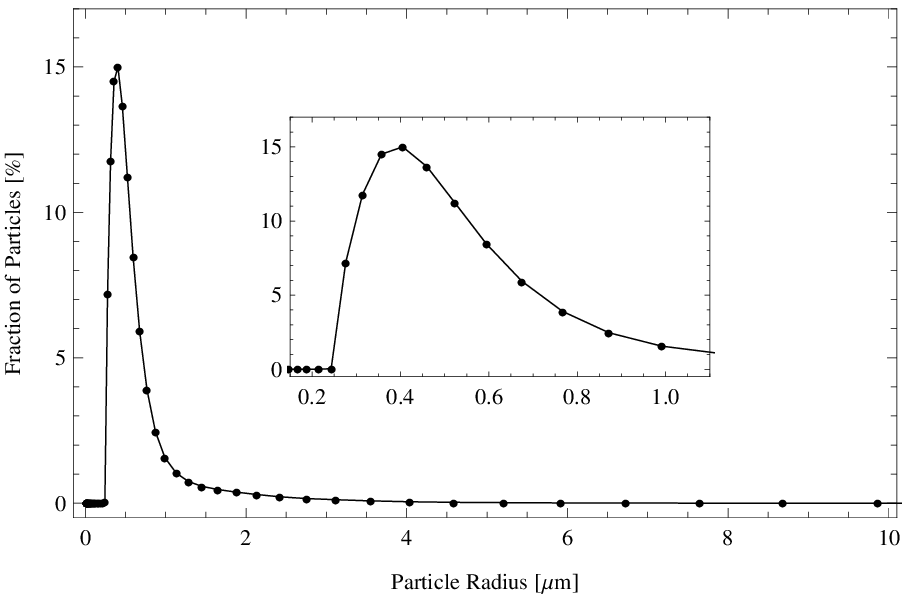}
\plotone{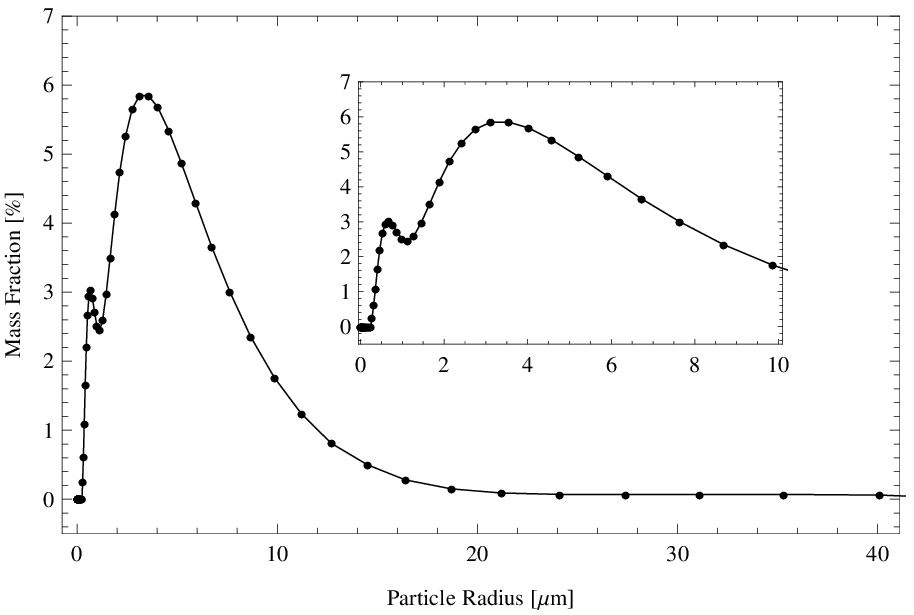}
\caption{Size and mass distribution of the Quartz dust grains used in the experiments: The plots show the fraction of the number of particles and the fraction of mass against the particle radius.\label{quarz}}
\end{figure}

The mechanical properties of silicate dust agglomerates primarily depend on two parameters, the size distribution of the dust grains and the aggregate porosity, as can be seen in a variety of experiments \citep{deckers2013, meisner2013, beitz2011, schraepler2012, blum2008}. As in our previous experiments \citep{deckers2013, meisner2012} we use quartz powder consisting of irregularly shaped grains (producer: Sigma-Aldrich). Fig. \ref{quarz} shows the size and mass distribution of the quartz dust measured with a particle size analyzer (Mastersizer 3000, manufacturer: Malvern Instruments). Here, the particle sizes are analyzed by laser diffraction, i.e by analyzing the scattered light of the dispersed particles. The mean particle radius from the size distribution is $0.45~\mathrm{\mu m}$ with a standard deviation of $0.13~\mathrm{\mu m}$. From the mass distribution we get a mean radius of $3.71~\mathrm{\mu m}$ and a standard deviation of $2.26~\mathrm{\mu m}$. 95\,\% of the particles are smaller than $1~\mathrm{\mu m}$, whereas around 63\,\% of the mass is in particles between $1$ and $6~\mathrm{\mu m}$. For irregularly shaped quartz powder of the same manufacturer, \citet{kothe2013} found mean radii of $0.63~\mathrm{\mu m}$ and $2.05~\mathrm{\mu m}$ for the size and mass distribution, respectively.\\

In this study we investigate the collisions of centimeter projectiles onto decimeter targets. The agglomerates are prepared in the same way as described by \citet{deckers2013} (decimeter targets) and \citet{meisner2012} (centimeter projectiles). The dust is compressed in a cylindrical mount with a hydraulic press and then pressed out of the mould. By doing so, we get cylindrical agglomerates (target bodies) with a diameter of 12cm, a height of around 12cm and a mean volume filling factor of $\Phi_{t,m}$$=0.44\pm$ 0.004.\\
The mass, being just over 1.5 kg, and height of every agglomerate are measured before the experiment in order to specify their volume and volume filling factor. Projectiles with three different diameters, 1.5\,cm, 2\,cm and 3\,cm, with masses between 2.5\,g and 20\,g and heights ranging from 1 to 3\,cm were used in the experiments. They were also compressed manually in a cylindrical mould to achieve a defined volume filling and geometry. By varying the mass of the projectiles, we can conduct experiments at different collision energies at similar collision velocities. Their mean volume filling factor is at $\Phi_{p,m}$$=0.466\pm0.02$. In the further analysis the difference in filling factor between projectile and target is defined as $\Delta\Phi = \Phi_p-\Phi_t$.\\

\subsection{Setup for the Collision Experiments}
\begin{figure}[ht]
\epsscale{1.}
\plotone{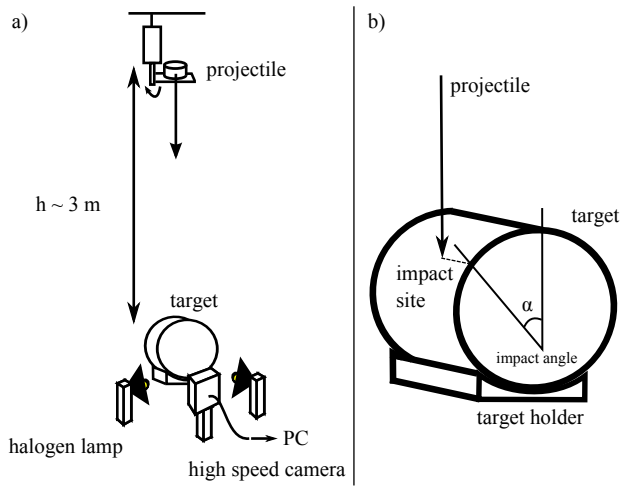}
\caption{a) Experimental setup for the collision of the centimeter projectile onto the decimeter target, b) Illustration of the impact angle $\alpha$. \label{aufbau}}
\end{figure}

\begin{figure}[hb]
\epsscale{1.}
\plotone{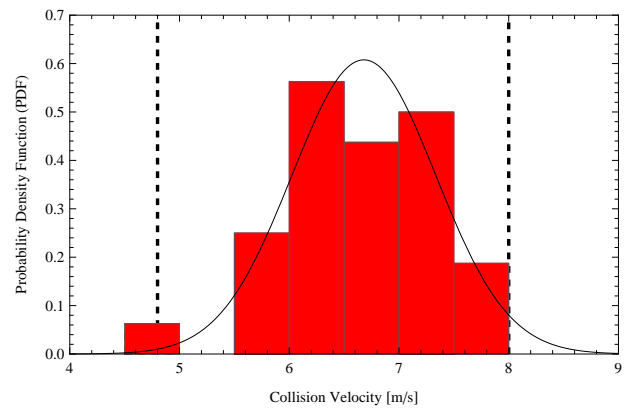}
\caption{Velocity distribution and distribution function (normal distribution, $\sigma$=0.67). The dashed lines show the lowest and the highest velocity, respectively. \label{gvert}}
\end{figure}

Both, the projectile as well as the target, are placed inside a vacuum chamber at typical ambient pressure of $p\lesssim5\cdot10^{-1}~\mathrm{mbar}$. The vacuum chamber is evacuated slowly, for about 1 hour, in order to prevent damage to target and projectile by the escaping gas. Collision experiments with dust agglomerates have to be carried out under vacuum conditions in order to neglect the influence of residual gas within the agglomerates and to reduce the influence of gas drag on the collision dynamics. The setup for the collision experiments is shown in Fig. \ref{aufbau}.\\
The projectile is placed onto an ejection mechanism, which mainly consists of a metal plate and a gear drive, at the top of the drop tube. When the evacuation of the tube and the vacuum chamber is complete, the gear drive moves the metal plate to the side and the projectile is dropped down. The sideward movement of the metal plate leads to a rotation of the projectile, which then drops down with an oblique orientation (see Fig. \ref{koll}). A high speed camera observes the collision with the cylindrical target at the bottom of the vacuum chamber at 500 frames per second. The chamber is illuminated by two halogen lamps.\\
The collision velocity is determined from the camera images. Due to variations in the ejection of the projectiles, the collision velocity varies, too. The mean collision velocity is $6.68\pm0.67~\mathrm{m\,s^{-1}}$. Fig. \ref{gvert} shows the velocity distribution. The black line is the Distribution Function, a normal distribution with a standard deviation of $0.67~\mathrm{m\,s^{-1}}$. The dashed lines show the lowest and the highest velocity, respectively.\\

\section{Results}

\begin{figure}[!ht]
\epsscale{1.}
\plotone{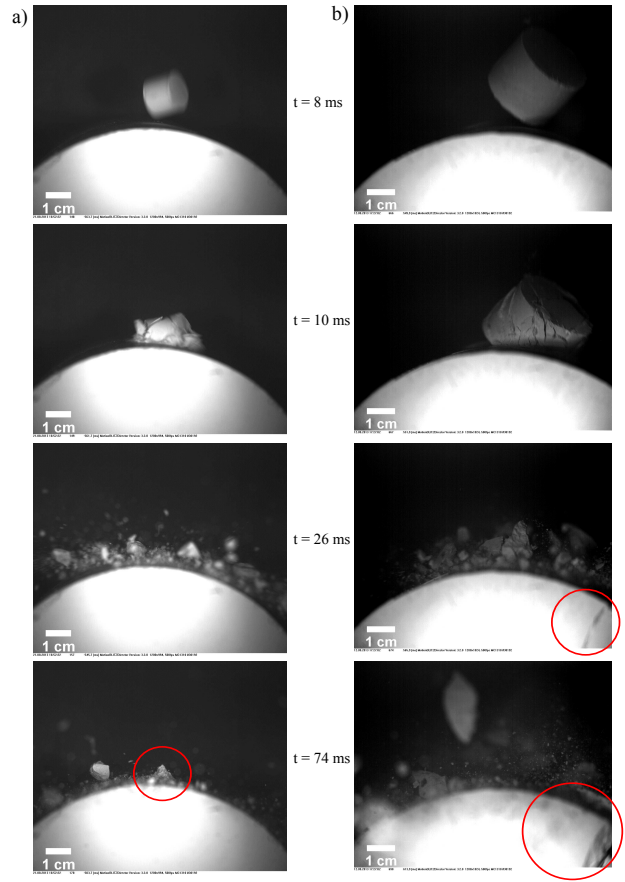}
\caption{Example for the two collision outcomes: a) mass gain, b) catastrophic disruption \label{koll} The circle in a) shows a typical grown dust cone, the circles in b) show typical target fragmentation. Most fragmenting lines are not visible, so the fragmentation strength is determined by measuring the mass of the biggest fragment.}
\end{figure}

The linear velocity of the projectile is determined by tracking the edges of the agglomerate. The agglomerates have a mean collision velocity of $6.68~\mathrm{m\,s^{-1}}$. Due to variations in the ejection mechanism the agglomerates follow the velocity distribution in Fig. \ref{gvert}. The variation in the ejection of the projectiles can also lead to a rotation around their symmetry and/or their transverse axis. For these cases the angular velocity is derived from the camera images as well. Together with the moments of inertia the rotational energy can be calculated. As both mass and velocity of the agglomerates vary, and thereby influence the outcome of a collision (see Fig. \ref{grenze}), and in order to include the rotational movement of an agglomerate when necessary, we take the collision energy of the agglomerates as reference for our results. The kinetic energy of the projectile is defined as $E_{\mathrm{kin}} = 1/2 \left( m v^2 + I_x \omega_x^2+I_y \omega_y^2\right)$, with $I_x$ and $I_y$ being the moment of inertia around the symmetry and transverse axis respectively, $\omega_x$ and $\omega_y$ the corresponding angular velocities and $m$ being the mass. It is important to note that the projectile rotation can only be derived from the two-dimensional projection of the agglomerate, as the experiments are observed only with one camera and no 3D-data are available. However, the contribution of the rotation energy is small in comparison to the total kinetic energy of the projectile. It is only significant in one of the collisions, where it makes up about one third of the kinetic energy.\\
Two different outcomes of collisions can be observed within this study. At lower collision energies only the projectile gets disrupted and a small part of it sticks to the target, so the target gains mass. The contact between the grown dust cone and the target material is firm, so the material does not drop off when the target is tilted or retrieved from the vacuum chamber. At higher collision energies catastrophic disruption of the target can be observed. Catastrophic disruption means, that the largest fragment has less than half the mass of the original agglomerate. There is a sharp transition between collisions with mass gain and catastrophic disruption of the target with no collision leading to only slight damage of the target. Table \ref{uebersicht} gives an overview of all experiments and their results.\\
Fig. \ref{koll} shows an example for both possible collision outcomes. The red circle in a) marks the grown structure on the target and in b) marks the area where the breaking up of the target is visible from the front. Similar to the fragmentation observed in mutual decimeter collisions \citep{deckers2013}, the fracture lines lie perpendicular to the symmetry axis of the agglomerate and are not well visible from the front.\\

\subsection{Threshold between mass gain and catastrophic disruption}

The method of error propagation was used to calculate the error of linear and angular velocity of an agglomerate, assuming an error of two pixels for the projectile position. Together with the errors in calculating the mass ($\Delta m=0.01\,\mathrm{g}$) and volume ($\Delta h=0.01\,\mathrm{cm}$) of the projectiles we can determine the error of the collision energy.\\

\onecolumn
\begin{table}[H]
\centering
\begin{tabular}{|c|c|c|c|c|c|c|c|c|} \hline
No.&$m_p~[\mathrm{g}]$&$\Delta\Phi$&$\alpha~[^{\mathrm{\circ}}]$&$v~[\mathrm{m s^{-1}}]$&$E~[\mathrm{mJ}]$&Result&$e_{ac}$[\%]&$M_F [\mathrm{g}]$ \\ \hline
1&2.53&0.059&0&7.54&71.92&0&3.95& \\ \hline
2&2.73&0.022&0&7.47&76.17&0&5.5& \\ \hline
3&2.62&0.047&34.5&6.18&50.03&0&4.73& \\ \hline
4&2.43&-0.06&11.8&6.57&52.45&0&7.24& \\ \hline
5&2.49&0.037&43.2&7.20&64.54&0&5.26& \\ \hline
6&2.57&0.013&5&6.92&61.53&0&6.89& \\ \hline
7&3.77&0.022&0&6.64&80.62&0&7.35& \\ \hline
8&2.61&0.046&20&6.25&51.04&0&5.21& \\ \hline
9&2.83&0.028&68&6.47&59.23&0&4.7& \\ \hline
10&11.5&0.015&0&6.52&244.14&0&12.3& \\ \hline
11&9.1&0.023&5&5.96&151.17&0&7.26& \\ \hline
12&5.44&0.029&38.9&6.31&109.54&0&5.63& \\ \hline
13&7.12&0.026&13.9&5.9&123.8&0&6.12& \\ \hline
14&7.46&0.039&35.2&6.47&156.33&0&4.4& \\ \hline
15&8.34&0.048&0&7.44&206.68&0&9.48& \\ \hline
16&9.03&0.036&32.8&6.62&197.89&0&7.51& \\ \hline
17&10.44&-0.029&24.5&7.46&290.03&0&11.23& \\ \hline
18&8.63&0.023&60.2&6.19&165.21&0&3.2& \\ \hline
19&9.98&0.012&22&7.51&294.66&0&9.97& \\ \hline
20&6.71&0.046&23&7.05&115.35&0&5.92& \\ \hline
21&6.98&0.036&48.5&4.81&129.75&0&3.65& \\ \hline
22&7.79&0.013&11.5&7.43&215.11&0&10.02& \\ \hline
23&7.19&0.008&58.1&7.26&189.23&0&4.98& \\ \hline
24&9.8&0.012&17&6.22&193.33&0&6.55& \\ \hline
25&10.39&0.02&19&5.99&186.19&0&9.24& \\ \hline
26&10.65&0.005&45&5.91&185.81&0&8.08& \\ \hline
27&10.69&0.022&69&6.8&246.86&0&6.11& \\ \hline
28&19.86&&20.7&6.35&400.15&1&&450 \\ \hline
29&16.26&0.025&7&6.43&335.82&1&&500 \\ \hline
30&14.28&0.016&29.2&6.6&311.4&1&&540 \\ \hline
31&11.98&0.074&0&7.31&320.35&1&&365 \\ \hline
32&9.44&0.039&32&7.99&301.1&1&&678 \\ \hline
\end{tabular}
\caption{Overview of the conducted experiments and their results. $m_p$ is the mass of the projectile, $\Delta\Phi$ the difference in filling factor ($\Delta\Phi = \Phi_p-\Phi_t$), $\alpha$ the impact angle, $v$ the collision velocity and $E$ the collision energy. The collision Result is set to 0 for mass gain and 1 for catastrophic disruption. $e_{ac}$ is the accretion efficiency in mass gain collisions and $M_F$ the mass of the largest fragment in fragmenting collisions.}
\label{uebersicht}
\end{table}

\twocolumn

\begin{figure}[ht!]
%\epsscale{1.}
\plotone{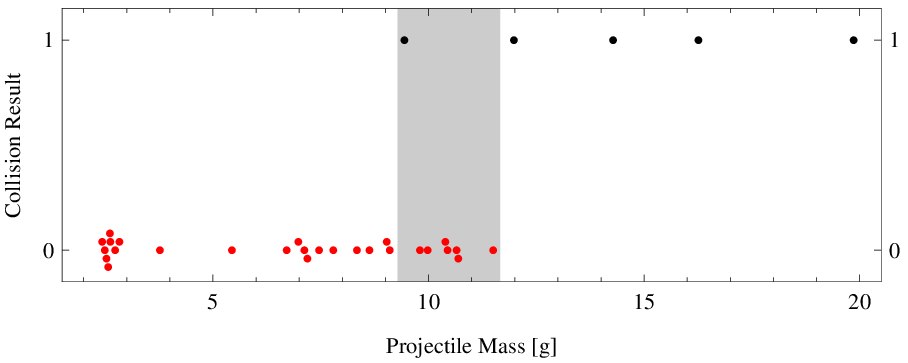}
\plotone{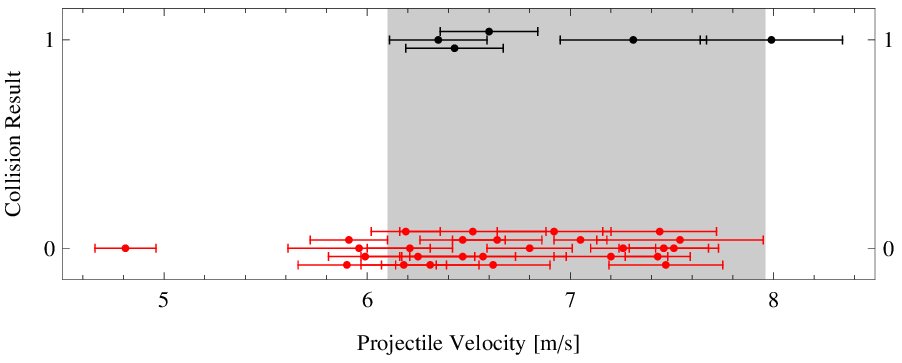}
\plotone{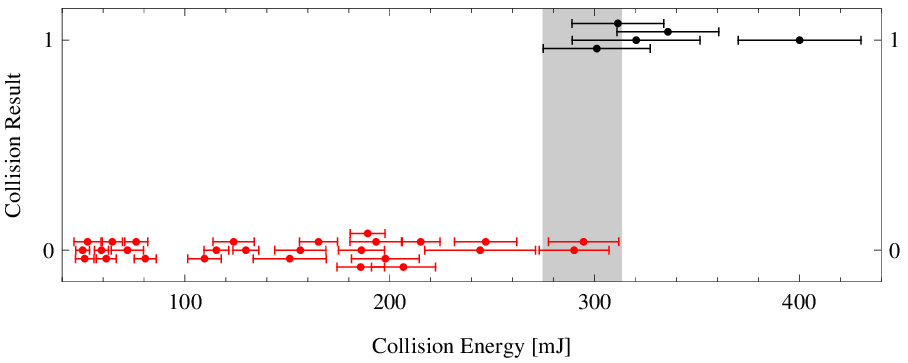}
\caption{Collision Result (0 = mass gain, 1 = catastrophic disruption) plotted against projectile mass, velocity and energy. The error in mass of \mbox{0.01}\,g is not shown in the plot. See text for details.\label{grenze}}
\end{figure}

Fig. \ref{grenze} shows the results of the collisions, with mass gain set to 0 and catastrophic disruption to 1, plotted against mass, velocity and collision energy of the projectiles. The error of the projectile mass is assumed to be \mbox{0.01}\,g (not shown in the plot). Data points are moved slightly on the y-axis for better visibility. The gray bars show the area between the highest value, where mass gain is observed and the lowest value for catastrophic disruption (including errors).\\

Variations both in mass and velocity influence the outcome of a collision. This can be seen in the top two plots, where the gray bars show the overlap in the two collision results. Therefore, neither mass nor velocity alone can be taken as independent parameter. Thus we take the collision energy as reference for the analysis of the collision experiments.\\
The bottom plot shows the collision result in dependency of the collision energy. The threshold between mass gain and catastrophic disruption is in the center of the area highlighted in gray. The threshold collision energy is $298\pm25~\mathrm{mJ}$.\\

\subsection{Accretion Efficiency}
\begin{figure}[ht]
\epsscale{1.}
\plotone{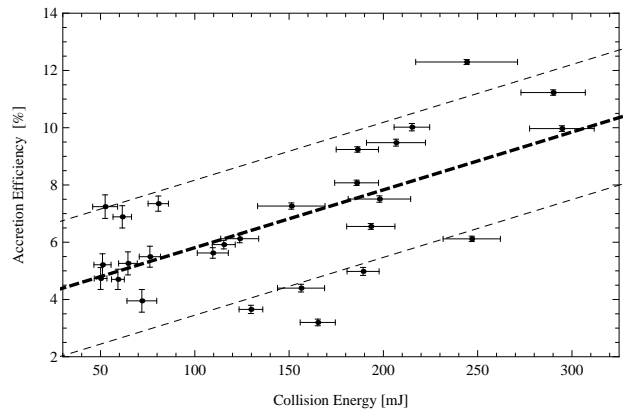}
\caption{Accretion Efficiency: the thick dashed line is a linear fit, the thin dashed lines show the range of results\label{akkretion}}
\end{figure}

After every collision with mass gain the target was retrieved from the vacuum chamber and the mass gain was determined by weighing the part of the projectile that sticks to the surface. Although the grown dust cone is firmly attached to the target surface, it can be removed from the surface by force. The accretion efficiency $e_{ac}$ can then easily be calculated by dividing the gained mass by the mass of the projectile prior to the collision. Once again the error is calculated by error propagation, assuming an error in the sticking mass of \mbox{0.01}\,g, which corresponds to about 1\% to 10\% of the sticking mass. Even at similar collision energies the accretion efficiencies vary quite a bit, shown by the two dashed lines in Fig. \ref{akkretion}. This is a result of the fact that the accretion efficiency does not only depend on the collision energy, but also on the volume filling factor and the impact angle, which is shown in the following sections.\\

\subsubsection{Dependency on the Impact Angle\label{angle}}

The impact angle $\alpha$ is determined by measuring the angle between the impact site of the projectile and the center of the cylinder mantle (see Fig. \ref{aufbau}). The rotation of the projectile and especially its orientation at the moment of impact can not be determined exactly, as only one camera is used. However, we consider the influence of the projectile orientation to be small and negligible in comparison to the impact angle resulting from the impact position with respect to the target.\\

\begin{figure}[hb]
\epsscale{1.}
\plotone{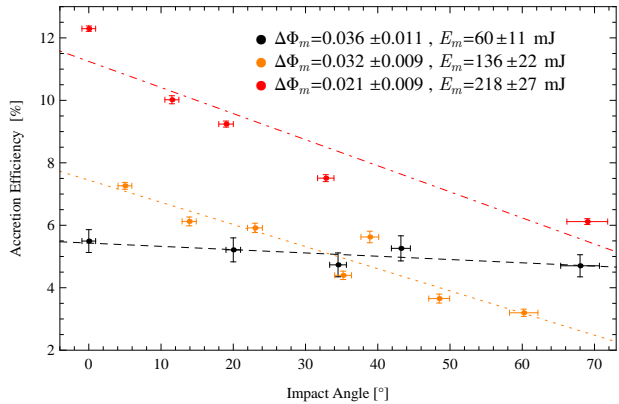}
\caption{Dependency of the Accretion Efficiency on the impact angle for collisions with similar collision energies and differences in filling factor. The lines in red, black and orange are linear fits to the data. \label{winkel}}
\end{figure}

In Fig. \ref {winkel} the accretion efficiency is plotted against the impact angle. The accretion efficiencies are divided into three groups. In every group the collisions have similar collision energies and similar differences in volume filling factor. Fig \ref{winkel} shows the mean values for these two parameters and their standard deviation. The residual scatter might be caused by the oblique orientation of the projectile, but this can not be quantified in more detail.\\
For low collision energies (black data points, dashed line) the accretion efficiency shows virtually no dependency on the impact angle, the linear fit follows the equation $5.43\,\mathrm{\%} - 0.01\,\mathrm{\%/degree}$. For medium (orange data points, dotted line) and higher energies (red data points, dotdashed line) a clear trend is visible that the accretion efficiency decreases with increasing impact angle. Here, the equation of the linear fit is $7.45\,\mathrm{\%} - 0.071\,\mathrm{\%/degree}$ and $11.25\,\mathrm{\%} - 0.084\,\mathrm{\%/degree}$, respectively. The accretion efficiency is highest for a central collision. For non-central collisions the accretion efficiency decreases with increasing impact angle.\\

\subsubsection{Dependency on the Volume Filling Factor\label{ff}}

As the collisions at low collision energies show (almost) no dependency on the impact angle, these results can be used to investigate the dependency on the difference in volume filling factor $\Delta\Phi = \Phi_p-\Phi_t$. Here, only collisions with similar collision energies are used.\\

\begin{figure}[hb]
\epsscale{1.}
\plotone{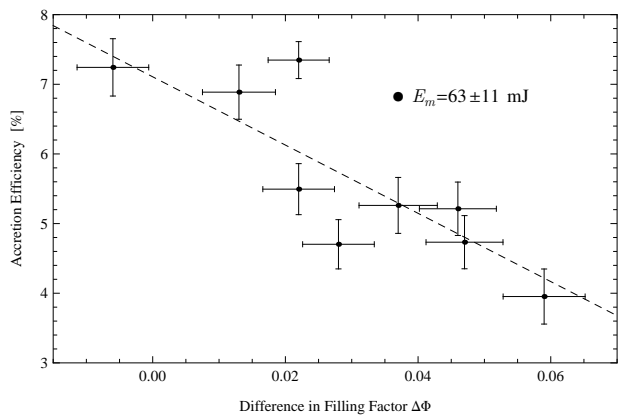}
\caption{Dependency of the Accretion Efficiency on the difference in filling factor for collisions with similar collision energies. The dashed line is a linear fit to the data. \label{fuell}}
\end{figure}

Fig. \ref{fuell} shows that the accretion efficiency decreases for an increasing difference in filling factor. The dashed line is a linear fit to the data with the equation $7.1\,\mathrm{\%} - (49 \cdot\Delta\Phi)\,\mathrm{\%}$. Here, the median collision energy is at $E_m= 63\pm11~\mathrm{mJ}$. This means that the more compact the projectile is in comparison to the target, the lower the accretion efficiency is. This trend can also be seen for projectiles that are more porous than the target, i.e. the difference in filling factor is negative.\\
\subsubsection{Normalized Accretion Efficiency}
Taking the dependency of the accretion efficiency on the impact angle (Fig. \ref{winkel}) and difference in volume filling factor (Fig. \ref{fuell}) into account, one can now take a closer look at the dependency on the collision energy. This can be done by including the slopes of the linear fits in Fig. \ref{winkel} and \ref{fuell} into the calculations: $e(E,\Delta\Phi=0,\alpha=0)=e(E,\Delta\Phi,\alpha)+(-49\cdot\Delta\Phi)+c_i\cdot\alpha$ (the values of $c_i$ are given below in Table \ref{variables}). Thereby we obtain the normalized accretion efficiencies, i.e. for a central collision of agglomerates of the same porosity.\\

\begin{figure}[ht]
\epsscale{1.}
\plotone{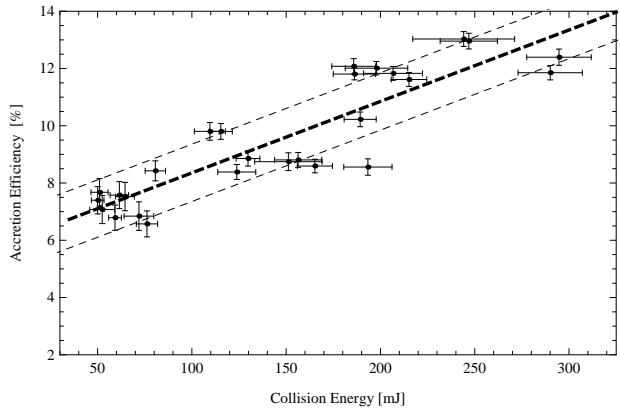}
\caption{Dependency of the normalized Accretion Efficiency on the collision energy. The thick dashed line is a linear fit to the data, the thin dashed lines show the rms error of the fit.\label{korr}}
\end{figure}

The normalized accretion efficiencies show a significantly reduced spread (see thin dashed lines in Fig. \ref{korr}). Moreover, the accretion efficiency increases with increasing collision energy. The thick black dashed line is a linear fit to the data with the equation $5.86\,\mathrm{\%} + 0.025\,\mathrm{\%/mJ}$ and a standard error of the slope of $0.003\,\mathrm{\%/mJ}$. The two thin dashed lines have the same slope and are obtained by adding and subtracting the root mean square (rms) error of the linear fit, respectively.\\

\subsection{Fragmentation Strength $\mu$}

\begin{figure}[ht]
\epsscale{1.}
\plotone{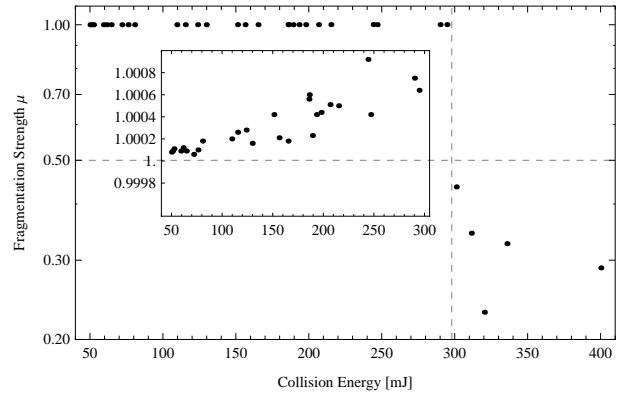}
\caption{Fragmentation Strength $\mu$ in dependency of the collision energy\label{mu}}
\end{figure}

After every fragmenting collision, the mass of the largest fragment $M_f$ of the target was weighed. The fragmentation strength $\mu ={M_f}/{M_0}$ can now be calculated by dividing the largest fragment mass by the mass of the target before the collision. Fig. \ref{mu} shows the values for $\mu$ plotted versus the collision energy.\\
For all fragmenting collisions the fragmentation strength is less than \mbox{0.5}. This shows that in all of these collisions we see catastrophic disruption of the target. In all other collisions $\mu$ is derived by dividing the target plus the sticking mass by the original target mass. Mass gain is very small in comparison to the target mass, so we get $\mu \cong1$ (see inset in Fig. \ref{mu}).\\

\subsection{Distribution of Projectile Fragments}
\begin{figure}[ht]
\epsscale{1.}
\plotone{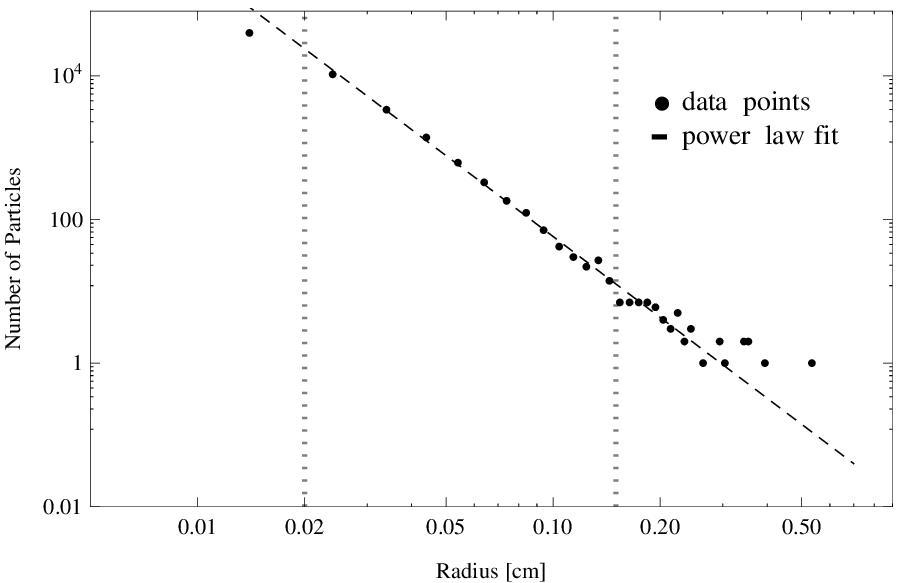}
\plotone{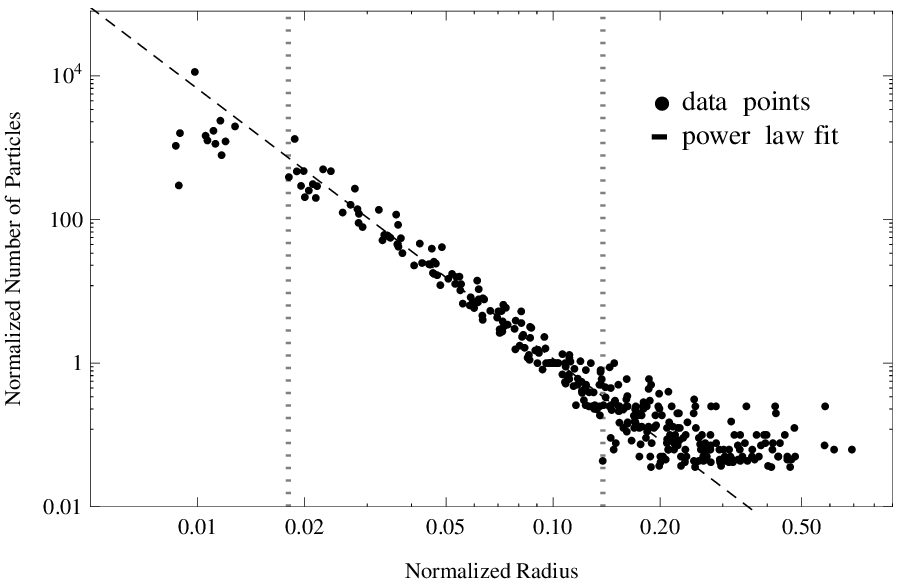}
\caption{Fragment distribution at a collision energy of \mbox{290}\,mJ and overview of all size distributions. The dotted lines mark the area where the distributions are fitted\label{vert}}
\end{figure}

In addition to the accretion efficiency the distribution of projectile fragments, in size as well as in mass, was analyzed as well. The smaller fragments were analyzed by placing them on a black background and taking camera images. From these images the cross section of every particle can be calculated as well as the radius assuming spherical particles. In this analysis we have a resolution limit of the cross section of five pixels. The biggest fragments were weighed and the radius was calculated from the mass, again under the assumption of spherical particles. The top part of Fig. \ref{vert} shows the distribution of particles for one collision (no. 17 in Table \ref{uebersicht}) at a high collision energy by plotting the number of particles as function of their radius (bin size 0.01\,cm). The dashed line is a power law fit with an exponent of $-3.7\pm0.1$, the dotted lines show the area where the distributions are fitted. All twelve analyzed particle distributions follow a power law, the average exponent is $-3.8\pm0.3$, the error is given by the standard deviation of all measurements. The lower part of Fig. \ref{vert} shows all size distributions and a power law with the exponent \mbox{-3.8}. Here, the normalized number of particles is plotted versus the normalized radius, i.e. the ratio of particle radius and radius of a sphere with the mass of the projectile. The number of particles is normalized to the value at a normalized radius of \mbox{0.1}\,cm for every distribution.\\
Besides the size distribution, we also analyzed the mass distribution of the projectile fragments. The mass of the small particles was calculated from their radius, assuming the particles are spherical and the volume filling factor remains unchanged in the collision.\\
The smallest particles, i.e. particles with radii of less than 0.1 cm, are by far the most numerous. As the mass of each of these particles is very small, their mass fraction is at a relatively low value. The few large particles however, make up a large fraction of the fragment mass, as their mass is much higher. In the analysis of the mass distributions we only took the distributions into account where the biggest particle has a radius of more than 0.4 cm. The other distributions with no such big fragment were neglected, as we can not rule out that a large fragment was disrupted in a secondary collision. Analysis of the ejecta velocities show that if big fragments move from the target freely, their average velocity is $2~\mathrm{m\,s^{-1}}$ (see circle 4 in Fig. \ref{ejecta}). This fast velocity of big fragments supports the idea that these fragments might break up in secondary collisions and in some size distributions no big fragment can be found. \citet{kothe2010} showed that millimeter aggregates fragment in collisions at velocities of a few $\mathrm{m\,s^{-1}}$. However, big fragments do not always move away from the target freely. In many collisions the biggest fragments roll along the surface of the target and are thereby not affected by secondary collisions at higher velocities. Whether secondary collisions influenced the smallest particles or not remains uncertain, as the fragment size distribution can not be derived from the camera images reliably (see Section \ref{fragvel}).\\

\begin{figure}[hb]
\epsscale{1.}
\plotone{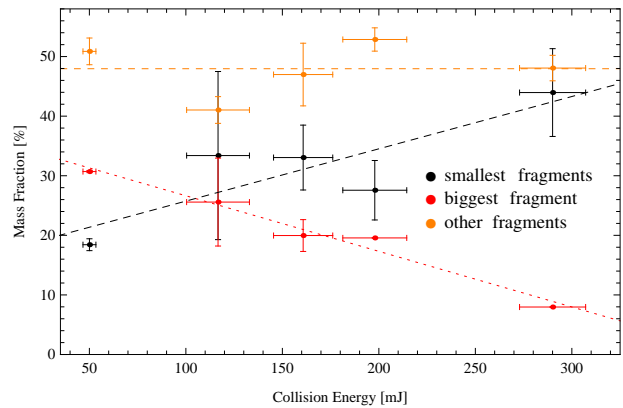}
\caption{ Averaged mass distribution of smallest fragments (radii $< 0.1~\mathrm{cm}$), the biggest fragment (radius $> 0.4~\mathrm{cm}$) and all other fragments with radii in between in dependency of the average collision energy.\label{vertm}}
\end{figure}

Fig. \ref{vertm} shows values for the mass fractions of the smallest particles, the biggest particle and all other fragments with radii in between of fragment distributions of collisions at different collision energies. The error was calculated taking an error of two pixels for the cross section of the small fragments and 5 mg for the weighed particles. For those mass fractions that were calculated by averaging the results of more than one distribution, the statistical error was added. The mass fraction of the smallest fragments by trend increases with incresing collision energy. On the other hand the fraction of the biggest fragment decreases.\\

\subsection{Projectile Fragment Velocities}\label{fragvel}

\begin{figure}[!ht]
\epsscale{0.9}
\plotone{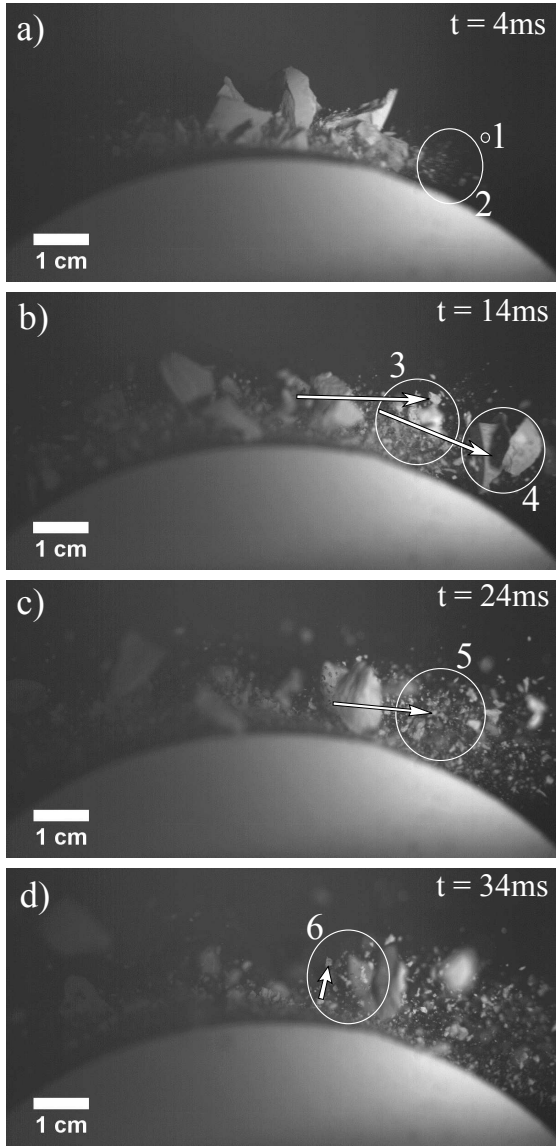}
\caption{Projectile Fragments ejected after a collision. The times given are the time differences to the collision moment. The arrows show the distance particles moved between two pictures.\label{ejecta}}
\end{figure}

\begin{figure}[ht]
\epsscale{1.}
\plotone{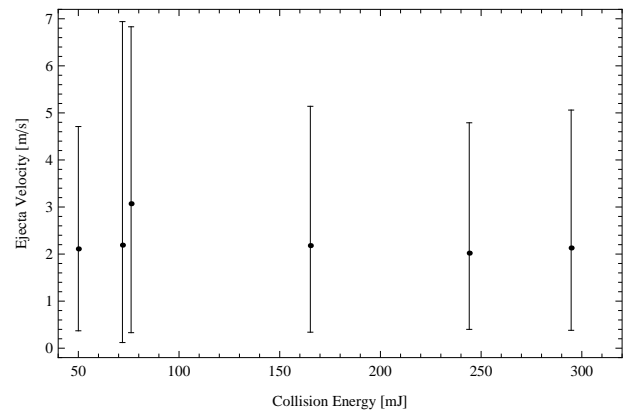}
\caption{Ejecta velocities at different collision energies: The data points give the mean ejecta velocity, the bars show the range of velocities.\label{ejvel}}
\end{figure}

Fig. \ref{ejecta} displays the break up of a projectile after a collision and the ejection of fragments. The time interval between pictures is \mbox{10}\,ms, the arrows show the distance particles moved between two pictures.\\
In the four pictures ejecta of different sizes and velocities can be seen. As there are a lot of ejecta and most of them move in groups close to one another it is difficult to analyze them. Even then it is possible to track individual fragments and thereby get the velocity not only of these individual particles but of the group of particles moving at the same velocity. In this way we can make a rough classification into groups of particles. These different groups of ejecta can be observed in all of the collisions.\\
Most of the ejecta move away from the impact site very fast, their velocities are between $1$ and $7~\mathrm{m\,s^{-1}}$ (circles 1,2,3 and 5). The fastest ejecta are therefore almost as fast as the projectiles before the collision, which have a mean velocity of $6.68~\mathrm{m\,s^{-1}}$.\\
Some ejecta however are significantly slower and move at average velocities of $0.3$ to $1~\mathrm{m\,s^{-1}}$ (circle 6). Most of these fragments follow a parabola and return to the target. In this case they do not stick to the target, but drop off the surface when the target is retrieved from the vacuum chamber. They therefore do not contribute to 
the measured accretion efficiencies.\\
Fig. \ref{ejvel} shows the mean ejecta velocity (data points) and the range of ejecta velocities (bars) for collisions at different collision energies. Both range and mean value do not change significantly with increasing collision energy. This is in good agreement with \citet{teiser2011a}, who find that for projectiles of constant mass the ejecta velocities do not change with increasing collision velocity.\\
The fastest ejecta with velocities of up to $7~\mathrm{m\,s^{-1}}$ are very small and therefore difficult to observe, especially when they are not in the focal plane of the camera. These ejecta where thus not found in all collisions.\\

\section{Discussion}
\subsection{Critical Fragmentation Strength}
In contrast to the previous experiments on mutual decimeter collisions \citep{deckers2013} the decimeter agglomerates fragment into several pieces independent of the collision point. For all fragmenting collisions the largest fragment had less than half the mass of the original agglomerate. This means that in these collisions we observed catastrophic disruption of the decimeter target. Although we do not observe a smooth transition to catastrophic disruption, we can constrain the value of $Q^*$ to an interval. The threshold collision energy of $298\pm25~\mathrm{mJ}$ corresponds to a critical fragmentation strength of $190\pm16~\mathrm{mJ\,kg^{-1}}$. This is a factor of four higher than the result for the threshold conditions for fragmentation in mutual decimeter collisions suggested \citep{deckers2013}. Gravity does not contribute to this larger value, as the target bodies are expected to be under constant tensile stress due to their own weight. As they are supported by a target mount which is much smaller than the target itself, gravity leads to a constant stress within the dust agglomerates. This means that in this study we measured the lower limit of the fragmentation strength.\\ 

\subsection{Accretion Efficiency and its dependecy on collision parameters}
The accretion efficiency depends on three collision parameters: the collision energy, the impact angle and the volume filling factor. The dependency on the impact angle can be seen in Fig. \ref{winkel}. The accretion efficiency is highest for central collisions and decreases with higher impact angle.\\
The dependency on the difference in volume filling factor can be seen in Fig. \ref{fuell}. The more porous the projectile is in comparison to the target, the higher the accretion efficiency becomes. This fits very well to the findings of \citet{beitz2011} who investigated the accretion efficiency for collisions of cylindrical dust agglomerates made up of the same quartz powder used in this study. They investigated mutual collisions of agglomerates with a diameter of 3 cm and mass of 13 g. The collision geometry used by \citet{beitz2011} is different from this study, as their projectiles were more porous than the targets, i.e. a negative difference in volume filling factor. They found that the accretion efficiency increases, the more porous the projectile is in comparison to the target.\\
Fig. \ref{korr} shows the dependency of the normalized accretion efficiency on the collision energy. The accretion efficiency increases with higher collision energies. This is in good agreement with the results of \citet{beitz2011} for mutual collisions of cm-sized cylindrical dust agglomerates as well. They found the accretion efficiency to be higher for increasing collision velocities and suggest a linear dependency. \citet{kothe2010} analyzed the accretion efficiency for multiple collisions of highly porous mm-sized dust projectiles ($\Phi=0.15\pm0.01$) onto cm-sized compact dust targets ($\Phi$\,$\approx0.4-0.5$) at collision velocities between $1.5~\mathrm{m\,s^{-1}}$ and $6~\mathrm{m\,s^{-1}}$. Their aggregates are made up of monodisperse, spherical $SiO_2$ particles with a diameter of $1.5~\mu \mathrm{m}$. \citet{kothe2010} also find an increase in accretion efficiency with increasing collision velocity. \citet{wurm2005} also analyzed collisions of mm-sized dust projectiles onto cm-sized dust targets (both have $\Phi$\,$\approx0.34$) made of irregular $SiO_2$-grains but found no clear dependency of the accretion efficiency for collision velocities between $6~\mathrm{m\,s^{-1}}$ and $15~\mathrm{m\,s^{-1}}$. For these velocities the accretion efficiency remains more or less constant around 10\,\%.\\
We are able to give an analytical function for the dependencies of the accretion efficiency $e_{ac}$ on the collision energy $E$, the difference in filling factor $\Delta\Phi = \Phi_p-\Phi_t$ and the impact angle $\alpha$:
\begin{equation}
e_{ac}(E,\Delta\Phi,\alpha)=e_{ac,0}+aE+b\Delta\Phi+c_i \alpha
\label{akkgleichung}
\end{equation}

where $e_{ac,0}=e_{ac}(\Delta\Phi=0)+e_{ac,i}(\alpha=0)+e_{ac}(E=0)$ is the offset in the accretion efficiency given by the y-intercepts in the linear fits in Fig. \ref{winkel}, \ref{fuell} and \ref{korr}. The values for the parameters in Eqn. \ref{akkgleichung} are given in Table \ref{variables}.\\
These results can easily be included into coagulation models as e.g. by \citet{drazkowska2013} or by \citet{windmark2012b}.\\
\begin{table}[ht!]
\centering
\begin{tabular}{|c|c|} \hline
a&$0.025~\mathrm{\%}/\mathrm{mJ}$ \\ \hline
b&-49~\% \\ \hline
$c_1$&$-0.01~\mathrm{\%/degree}$\\
$e_{ac,1}(\alpha=0)$&5.43~\%\\
&$E\lesssim80\mathrm{mJ}$\\ \hline
$c_2$&$-0.071~\mathrm{\%/degree}$\\
$e_{ac,2}(\alpha=0)$&7.45~\%\\
&$80 \mathrm{mJ}\lesssim E\lesssim170\mathrm{mJ}$ \\ \hline
$c_3$&$-0.084~\mathrm{\%/degree}$\\
$e_{ac,3}(\alpha=0)$&11.25~\%\\
&$170 \mathrm{mJ}\lesssim E\lesssim250\mathrm{mJ}$\\ \hline
$e_{ac}(\Delta\Phi=0)$&7.1~\% \\ \hline
$e_{ac}(E=0)$&5.86~\% \\ \hline
\end{tabular}
\caption{Values for the parameters in Eqn. \ref{akkgleichung}}
\label{variables}
\end{table}\\

\subsection{Fragment Distribution and Velocity}

The distributions of the fragment sizes and velocities are critical parameters for coagulation models. Collisions as analyzed here are triggered by the different motion of particles in the gaseous environment of the disk. The projectiles are still well (or better) coupled to the gas, while the target bodies already drift fast. The targets are therefore exposed to a constant head wind, which is of the same magnitude as the relative velocity between target and projectile. All impact ejecta are therefore exposed to the same head wind, which accelerates these ejecta back toward the target. This mechanism has also been discussed by \citet{teiser2009} and \citet{ wurm2001b, wurm2001a}. If the ejecta are slow enough this acceleration can be sufficient to drive them back to the target body and increase the accretion rate significantly. In their simulations \citet{sellentin2013} do not find re-accretion for impacts of small projectiles onto larger targets in the viscous flow regime, where the mean free path is much smaller than the particle. \citet{sellentin2013} find re-accretion only in the free molecular flow regime, where the mean free path is much bigger than the particle. However, their study does not investigate whether re-accretion is possible in the regime between viscous and free molecular flow.\\
The projectile fragment distributions all follow a power law (see Fig. \ref{vert}) with an average exponent of \mbox{-3.8}. This power law size distribution is in good agreement with other collision experiments as well as simulations and analytical calculations of collision cascades involving secondary collisions. \citet{dohnanyi1969} developed a model to describe the collisional evolution within the asteroid belt and proposed a power law size distribution with exponent -3.5. \citet{pan2012} analyzed collision cascades analytically as well as in simulations and found power law size distributions with exponents between -3 and -4. Collision experiments with targets of different materials, like mortar or silicates, at different impact velocities find power laws with exponents in the same range \citep{davis1990, takasawa2011}. The size distribution also fits to simulations of \citet{geretshauser2011a}, who propose a four-population model for collision fragments, with a sub-resolution population, a power law population and two big fragments. \citet{paszun2009} simulated collisions of small highly porous dust agglomerates, with a diameter of $500\,\mathrm{\mu m}$ and $\Phi=0.16 - 0.25$, resulting in power law size distributions with exponents between \mbox{-0.3} an \mbox{-1.7}. These simulations might indicate that the size distribution without a collision cascade follows a shallower power law. Experiments by \citet{blum1993} on mutual collisions of $ZrSiO_4$-aggregates with sizes of \mbox{0.2}\,mm to \mbox{5}\,mm at a collision velocity of about $4~\mathrm{m\,s^{-1}}$, without secondary collisions, show a power law size distribution with exponent \mbox{-3.38}. This is close to the exponent found in this study and suggests that secondary collisions might not influence the size distribution significantly.\\
The size distribution is important in various astrophysical environments where fragmenting collisions play an important role, e.g. debris disks \citep{krivov2000} or the Kuiper and the asteroid belt in the solar system.\\
The distribution of projectile fragments in Fig. \ref{vertm} shows the trend, that the mass fraction of smallest fragments, i.e. particles with radii less than \mbox{0.1}~cm, increases with increasing collision energy. Due to the density of the cloud of ejecta we can only make a rough classification of ejecta velocities. In all analyzed collisions we find the same groups of ejecta. Furthermore, the mean ejecta velocity and the velocity ranges of ejecta do not change significantly with increasing collision energy. This suggests that the ejecta velocities do not increase at higher collision energies and that re-accretion is more likely for higher collision energies. In addition to that, the accretion efficiency increases with collision energy as well, leading to an enhanced growth of the bigger collision partner at higher collision energies. However, it is unclear how secondary collisions influenced the size and mass distribution.\\
The analysis of the projectile ejecta (see Section \ref{fragvel}) shows that most of the medium and small particles move away from the impact site at quite high velocities of several $\mathrm{m\,s^{-1}}$. These ejecta are too fast to be re-accreted by gas drag and remain as small particles in the protoplanetary disk. The group of slowest fragments however get ejected at velocities of \mbox{0.3} to $1~\mathrm{m\,s^{-1}}$. These fragments have to be considered if re-accretion by gas drag is analyzed.\\
The collisions of centimeter and decimeter agglomerates at around $7~\mathrm{m\,s^{-1}}$ analyzed in our experiments fit well to the assumed relative velocities between bodies of these sizes at 1 AU in protoplanetary disks \citep{weidenschilling1993}. Assuming the minimum mass solar nebula (MMSN, \citet{hayashi1985}) as disk model, the mean free path of gas molecules at 1 AU distance is at around 6 cm. This means that mean free path and target size are of the same order of magnitude and that we are in the regime between free molecular flow and viscous flow. In their experiments \citet{wurm2001a,wurm2001b} showed that re-accretion is still possible if the larger body is ten times larger than the mean free path.\\
The slow ejecta observed in our experiments land back on the target due to gravity and follow parabolic trajectories similar to those shown by \citet{wurm2001a}. The acceleration by gas drag is given by $a = v_e/\tau$, where $v_e$ is the velocity of ejecta relative to the head wind and $\tau$ the coupling time of particles to the gas \citep{teiser2009}. In the free molecular flow regime $\tau$ is given as $\tau = 3.28\cdot10^5\frac{1}{\mathrm{m}}\,r\,\left(\frac{R}{1~\mathrm{AU}}\right)^3\,s$ \citep{blum1996,teiser2009}. The size distribution shows that most of the projectile fragments have radii of about $0.1\,\mathrm{mm}$. For a particle with radius $0.1\mathrm{mm}$ at 1 AU we get $\tau \approx 32\mathrm{s}$. Most of the slow fragments are ejected against the incomming headwind (see Fig. \ref{ejecta}). For these particles the acceleration by gas drag is enough to return them to the target if the ejecta velocity is significantly slower than the headwind of $7~\mathrm{m\,s^{-1}}$ and not more than about $0.5~\mathrm{m\,s^{-1}}$. In the analysis of the ejecta velocities we find that a significant number of slow ejecta have velocities between \mbox{0.3} and $0.5~\mathrm{m\,s^{-1}}$. This means that small and slow ejecta can be re-accreted to the target. A more detailed analysis of re-accretion and its efficiency is beyond the scope of this study, but is subject to current work by \citet{jankowski2014}.\\ 

\section{Conclusions}
Two different collision outcomes were observed for the collisions of centimeter and decimeter agglomerates, made up of irregular micro-meter quartz grains having a volume filling factor of $0.466\pm0.02$ and $0.44\pm0.004$ respectively: mass gain and catastrophic disruption of the decimeter target. From the analysis of the threshold condition for the catastrophic disruption of the decimeter agglomerates we can constrain the critical fragmentation strength to an interval, i.e. $Q^*=190\pm16~\mathrm{mJ\,kg^{-1}}$.\\
At lower collision energies the target gains a small amount, 3 to 12\,\%,  of the projectiles mass. The accretion efficiency depends on the collision energy and other collision parameters, impact angle as well as aggregate porosity (see Eq. \ref{akkgleichung}). With increasing collision energy the accretion efficiency becomes higher.\\
Additionally the analysis of projectile fragments shows a power law size distribution, possibly a result of a collision cascade involving secondary collisions. The exponent of $-3.8\pm0.3$ is in good agreement with simulations as well as other experiments.\\
Small projectile ejecta (radii $<0.1\,\mathrm{mm}$) can return to the target and be re-accreted to the decimeter agglomerate, if they are significantly slower than the headwind. Analysis of the mass distribution reveals an increase in smallest fragments for higher collision energies. This leads to a higher re-accretion probability for collisions of bodies of different sizes within protoplanetary disks and thereby increased growth of the bigger body, as it is these smallest fragments that can be re-accreted.\\

\acknowledgments

We would like to thank the Deutsche Forschungsgemeinschaft (DFG) for their funding within the frame of the SPP 1385 "The first 10 Million Years of the Solar System -- A Planetary Materials Approach".\\
We also thank the anonymous referee for his detailed and constructive comments, that helped to improve this paper. 
%\clearpage

\bibliography{literature}

\end{document}